\documentclass[journal]{IEEEtran}
\ifCLASSINFOpdf
\else
\fi
\usepackage{graphicx}
\usepackage{algorithm}
\usepackage{algorithmic}
\usepackage[pagebackref=false,breaklinks=true,letterpaper=true,colorlinks,urlcolor=blue,citecolor=blue,linkcolor=blue,bookmarks=false]{hyperref}
\usepackage[numbers,sort]{natbib}
\usepackage{amsmath}
\usepackage{makecell}
\usepackage{multirow}
\usepackage{wasysym}
\begin{document}
\newcommand{\figdir}{figures}

%
\title{Multimodal Topic Learning for Video Recommendation}
%
%
%


\author{Shi Pu$^*$ \ \ \ Yijiang He$^*$ \ \ \ Zheng Li \ \ \ Mao Zheng\\
        \textbf{Tencent AI Lab}
        \thanks{$^*$Equal Contribution}
        }

%
%


\markboth{IEEE Transactions on Image Processing}{}

%



\maketitle

\begin{abstract}

Facilitated by deep neural networks, video recommendation systems have made significant advances.
Existing video recommendation systems directly exploit features from different modalities (e.g., user personal data, user behavior data, video titles, video tags, and visual contents) to input deep neural networks, while expecting the networks to online mine user-preferred topics implicitly from these features.
However,  the features lacking semantic topic information limits accurate recommendation generation.
In addition, feature crosses using visual content features generate high dimensionality features that heavily downgrade the online computational efficiency of networks. 
In this paper, we explicitly separate topic generation from recommendation generation, propose a multimodal topic learning algorithm to exploit three modalities (i.e., tags, titles, and cover images) for generating video topics offline.
The topics generated by the proposed algorithm serve as semantic topic features to facilitate preference scope determination and recommendation generation.
Furthermore, we use the semantic topic features instead of visual content features to effectively reduce online computational cost.
Our proposed algorithm has been deployed in the Kuaibao information streaming platform.
Online and offline evaluation results show that our proposed algorithm performs favorably.

%
%
%
%
%
%
%
%
%
%
\end{abstract}

\begin{IEEEkeywords}
Recommendation, video topics, multimodality.
\end{IEEEkeywords}

%
\IEEEpeerreviewmaketitle

\section{Introduction}
With great commercial and academic value, video recommendation has received growing attention in recent years with various applications.
One of the popular applications is the Kuaibao information streaming platform which utilizes a video recommendation system based on deep learning to attract tens of millions of users.
Existing video recommendation systems \cite{covington2016deep} typically consist of the candidate generation module and the ranking module.
When recommending videos to a user, the candidate generation module firstly retrieves hundreds of videos as candidates from a large-scale video pool.
Then, the ranking module assigns scores to these candidates.
A few videos with the highest scores are ranked by the scores and shown to the user.
These two modules make decisions according to the features which are generated by feature crosses between features from users and features from videos.
Existing deep video recommendation systems \cite{ge2018image} have demonstrated great success by exploit features from different modalities (e.g., user personal data, user behavior data, video titles, video tags, and visual contents). 

The deep networks in existing video recommendation systems are expected to online mine user-preferred topics implicitly from input features for preference scope determination.
However, on the one hand, video topics are abstracted from the contents of a large number of videos in the video pool, and there are no definite classes for video topics.
On the other hand, existing approaches use the deep networks pre-trained on recognization datasets to generate features from video contents (i.e.,  video titles, video tags, and visual contents).
Thus, existing video content features can not reflect on semantic topic information as these features lack global information related to the video pool.
Although using features from video contents is critical for helping understand the content preferences of users, existing video content features without semantic topic information limit accurate recommendation generation.
In addition, limited by computational resources, existing deep video recommendation systems mainly use convolutional neural networks (CNNs) to extract features from cover images of videos as the visual content representations.
However, using CNN features for feature crosses still generates high dimensionality features which heavily downgrade the online computational efficiency of the deep networks in video recommendation systems. 
It is therefore of great importance to learn semantic topic features with low dimensionality from video contents to replace existing CNN features.

\renewcommand{\tabcolsep}{.8pt}
\def\swone{0.9\linewidth}
\begin{figure}[t]
\begin{center}
\begin{tabular}{c}
\includegraphics[width=\swone]{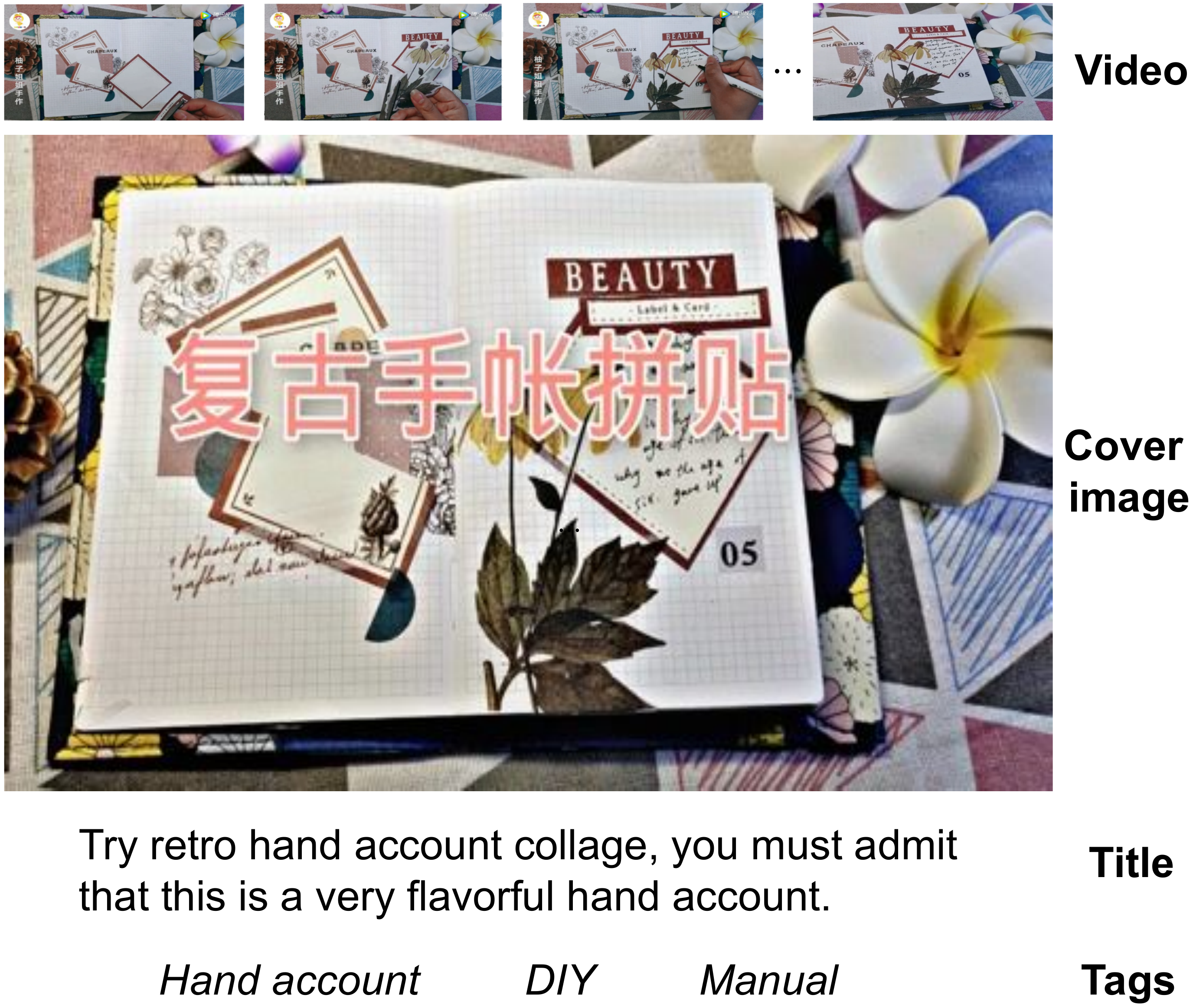}
\end{tabular}
\end{center}
\caption{Visualization of video content including tags, title, and cover images.}
\label{fig:fig1}
\end{figure}

In this paper, we propose a multimodal topic learning (MTL) algorithm which uses three modalities (i.e., tags, titles, and cover images) to generate video topics offline.
Figure \ref{fig:fig1} shows a video with these three modalities.
The titles and cover images describe the video contents in general terms from the perspective of video publishers, and the tags describe the video contents more objectively and fine-grained.
Our MTL algorithm explicitly learns semantic topics from these three modalities to facilitate preference scope determination and recommendation generation.
In this algorithm, we firstly exploit a multi-task learning method to end-to-end learn deep features from cover images and titles.
Meanwhile, a graph embedding approach is proposed to generate tag features.
Then, we combine the deep features with the tag features as topic representations for each video.
%
%
%
We collect a large-scale set of topic representations of videos from the video pool and use a topic clustering process to handle this set for semantics topics generation.
After that, we assign semantic topics for each video according to their topic representations.
The semantic topics of each video are used to improve existing video content features for accurate recommendation generation while using the low dimensionality topic features instead of CNN features effectively reduce online computational cost.
Our proposed MTL algorithm has been adopted by the Kuaibao information streaming platform. 
Through extensive online and offline evaluations, we demonstrate that the proposed MTL algorithm performs favorably.

The main contributions of this work are:
\begin{itemize}
\item We propose a multimodal topic learning algorithm to generate semantic topics.
\item We use the topics as semantic topic features to advance existing video recommendation systems.
\item We conduct extensive online and offline evaluations. The results show that the proposed algorithm performs favorably.
\end{itemize}

\section{Related work}
Recommendation systems have been widely discussed in the literature \cite{davidson2010youtube}. In this section, we mainly discuss the representative recommendation algorithms, and multimodal learning most relevant to this work. 
\flushleft\textbf{Recommendation algorithms.}
Historical behavior records of users build a binary relationship between users and items.
Recommendation algorithms use this relationship to recommend items that the users may be interested in.
Early recommendation algorithms \cite{koren2008factorization} exploit shallow models to model this relationship. 
For example, the Collaborative Filtering (CF) method \cite{sarwar2001item} exploits the similarities among users or items for recommendation.
The FM algorithm \cite{rendle2010factorization,juan2016field} uses factorized parameters to perform second-order feature interactions.
With the development of deep learning, existing recommendation algorithms have demonstrated great success by exploiting deep features \cite{devlin2018bert,szegedy2016inception,chollet2017xception} and decision-makers \cite{cheng2016wide,guo2017deepfm,wang2017deep,zhou2018deep}.
The Wide\&Deep model \cite{cheng2016wide} uses the memorization of the wide module and the generalization of the deep module to balance the accuracy and expansibility of recommendation systems.
Since the Wide\&Deep model, numerous methods have been developed to improve this framework for recommendation systems.
The DeepFM approach \cite{guo2017deepfm} proposes to utilize the FM layer instead of the Wide module to improve the Wide\&Deep model.
The DCN method \cite{wang2017deep} uses the Cross network to replace the Wide module in the Wide\&Deep model for high order feature interactions. 
The DIN algorithm \cite{zhou2018deep} integrates an attention mechanism for feature selection into deep recommendation systems.
Existing state-of-the-art deep video recommendation algorithms \cite{davidson2010youtube} focus on capturing video content features to understand user preferences. 
These video content features are mainly from the titles \cite{peters2018deep,devlin2018bert} and cover images \cite{simonyan2014very,he2016identity,szegedy2016rethinking,szegedy2016inception,chollet2017xception} of videos.
However, existing methods pay less attention to the global semantic topics as video content features.
Our proposed algorithm assigns topics to each video to facilitate preference scope determination and recommendation generation.

\flushleft\textbf{Multimodal learning.}
Multimodal learning aims to achieve the ability to process and understand multi-source modality information include images, videos, audios, and semantics.
The multimodal learning scheme has been widely exploited for many applications, include machine translation \cite{afouras2018deep}, image captioning \cite{yao2018exploring}, video captioning \cite{pan2017video}, speech synthesis \cite{sotelo2017char2wav}, image semantic segmentation, mobile identity authentication, etc.
In video recommendation, existing methods \cite{huang2019multimodal} directly use video content features from different modalities for recommendation decisions.
The redundancy between modalities downgrades the recommendation performance.
In this work, we propose a multimodal multitask learning approach to learn joint representations from the titles and cover images. 
Our approach aims to emphasize the complementarity and synchrony between modalities to represent and summarize the multimodal data. 
\section{Proposed method}

\renewcommand{\tabcolsep}{.8pt}
\def\swone{0.96\linewidth}
\begin{figure*}[t]
\begin{center}
\begin{tabular}{c}
\includegraphics[width=\swone]{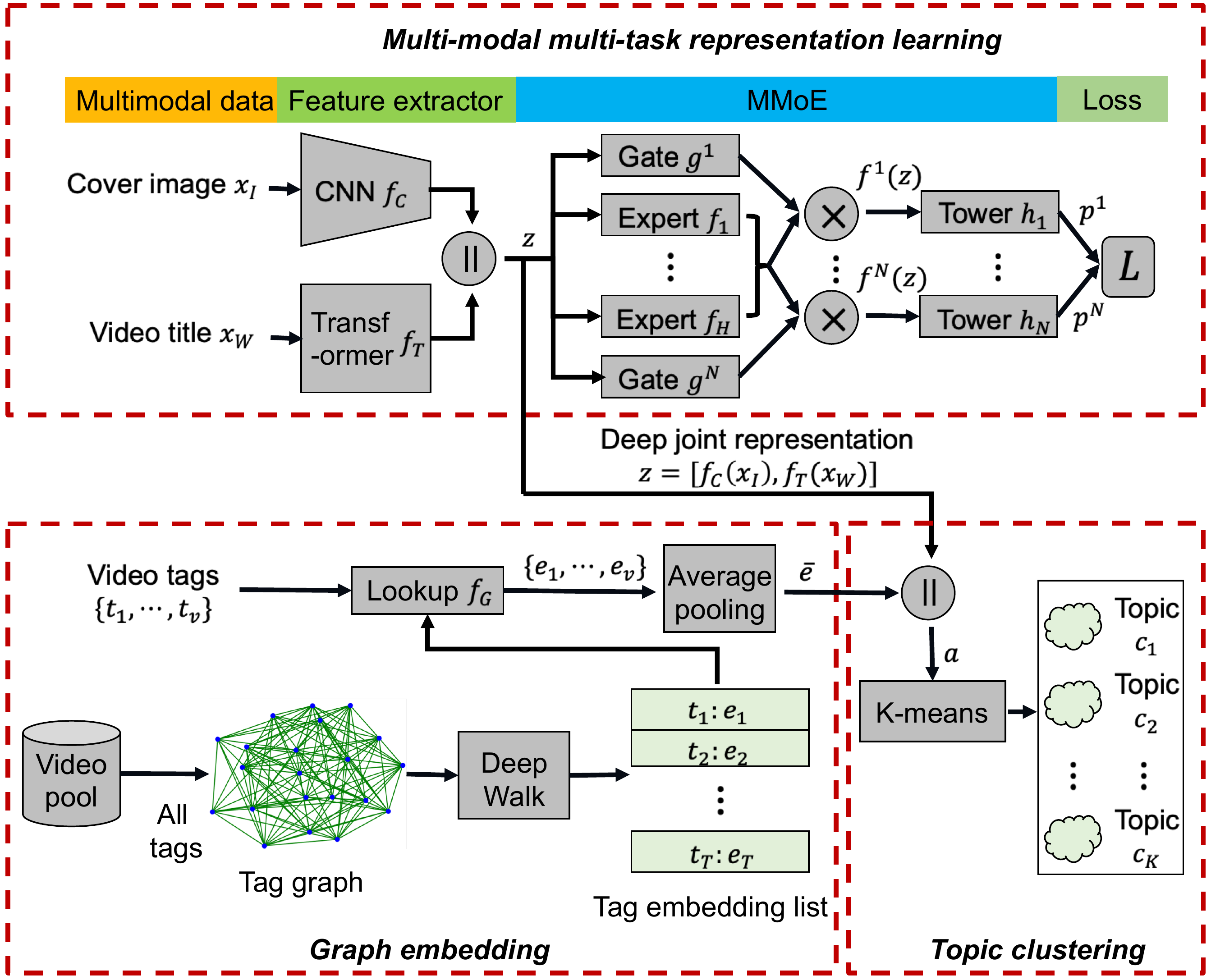}
\end{tabular}
\end{center}
\caption{Overview of the proposed multimodal topic learning algorithm. This algorithm consists of a multimodal multitask representation learning method, a graph embedding approach, and a clustering process. We use this algorithm to generate topics from multimodel data of input videos. For presentation clarity, $\times$ is $\sum_{i=1}^Hg^k(z)_if_i(z)$, $k \in\{1,\cdots,N\}$. $||$ is the concatenation operation.}
\label{fig:fig2}
\end{figure*}
\renewcommand{\tabcolsep}{.8pt}
\def\swone{0.96\linewidth}
\begin{figure*}[h]
\begin{center}
\begin{tabular}{c}
\includegraphics[width=\swone]{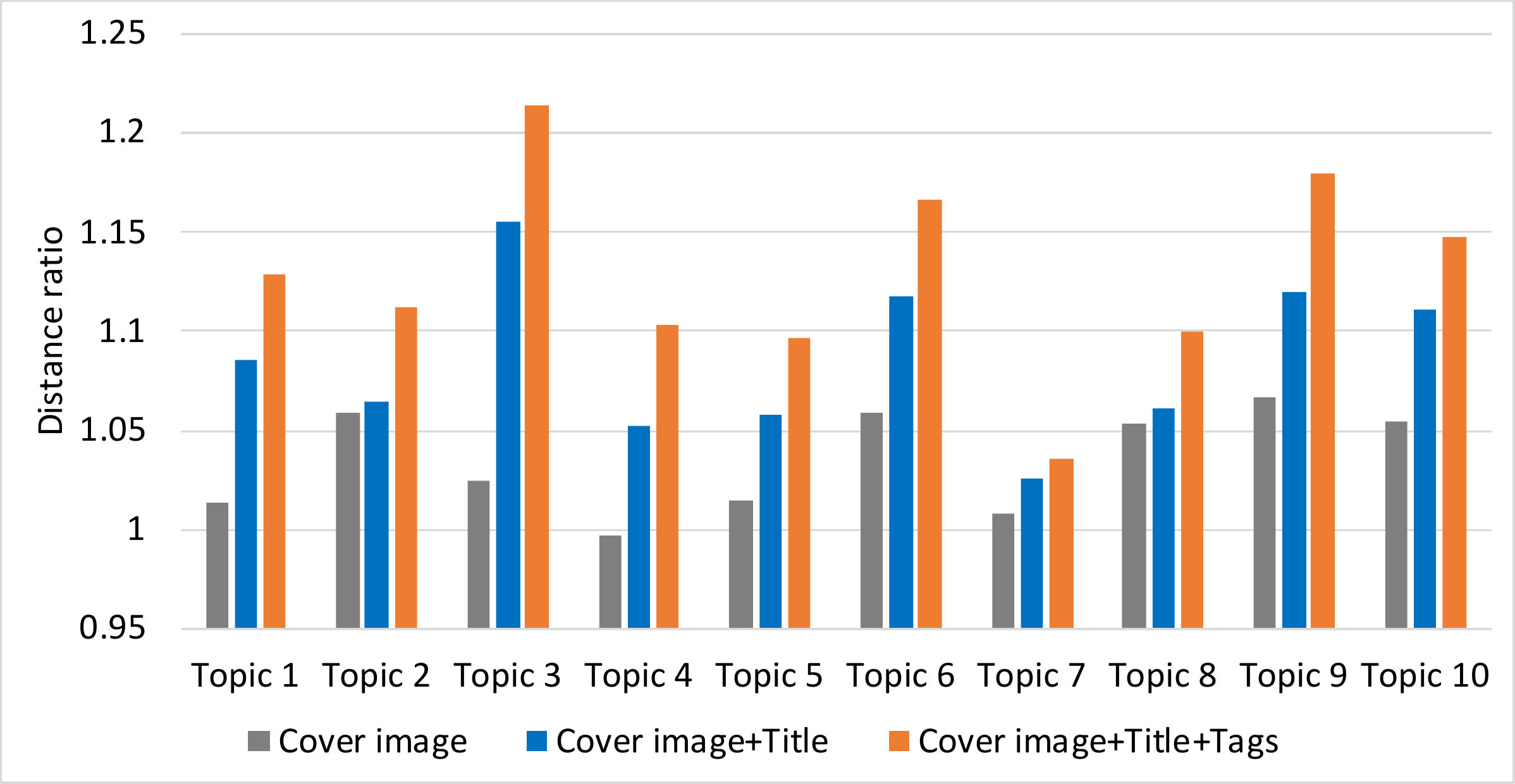}
\end{tabular}
\end{center}
\caption{Comparisons of distance ratios between average inter-class distances and average intra-class distances in 10 topic clustering results.}
\label{fig:fig3}
\end{figure*}
In this section, we first present how our multimodal topic learning algorithm generate topics, then we illustrate how the topics advance the candidate generation module and the ranking module in our video recommendation system.

\subsection{Multimodal topic learning}
As shown in Figure \ref{fig:fig2}, the multimodal topic learning scheme consists of a multimodal multitask representation learning method to generate deep joint representations from the titles and cover images, a graph embedding approach for tag representation generation, and a clustering process to generate topics based on the concatenate representations of the deep joint representations and the tag representations.
In the following, we introduce these three processes, respectively.
\subsubsection{Multi-modal multi-task representation learning}
Our network for the multimodal multitask representation learning method contains a feature extractor and a Multi-gate Mixture-of-Experts (MMoE) module \cite{ma2018modeling}.
The feature extractor consists of a pre-trained CNN module $f_C$ and a pre-trained  transformer module $f_T$ to generate deep features from the cover image $x_I$ and the title consisting of words $x_W$, respectively.
$x_I$ and $x_W$ are from an input video.
A straightforward method to generate the joint representations from the title and the cover image is concatenating the deep features $f_C(x_I)$ and $f_T(x_W)$.
However, the redundancy between different modalities will downgrade the recommendation performance.
We perform end-to-end multimodal representation learning to ensure the deep joint representations reflect the complementarity and synchrony between modalities to represent and summarize the multimodal data.
To improve the generalization ability of the deep joint representations, we use a multitask learning strategy MMoE to perform multimodal representation learning.
Compared to the early multitask learning methods that share bottom layers, the MMoE framework uses $H$ expert networks coupled with $N$ gating networks to generate representations for different $N$ tasks.
This strategy is effective to alleviate the task conflict problem.
The MMoE strategy can be denoted as:
\begin{equation}
\begin{split}
p^k=&h^k(f^k(z)),\\
f^k(z)=&\sum_{i=1}^Hg^k(z)_if_i(z),\\
z=&[f_C(x_I), \ f_T(x_W)],\\
k \in&\{1,\cdots,N\},
\label{fm:fm1}
\end{split}
\end{equation}
where $z$ is a input feature that is a concatenate representation of $f_C(x_I)$ and $f_T(x_W)$.
$f_i$ is an expert network. 
$g^k$ is a gating network with the softmax function which ensures $\sum_{i=1}^Hg^k(z)_i=1$ and $g^k(z)_i>0$.
$h^k$ is a tower network that generates the output $p^k$ corresponding to the $k$-th task.
In our method, $p^k$ is a prediction probability distribution that the input video belongs to the preset classes.
$N$ is 3 as we perform 3 classification tasks to learn the deep joint representation $z$.
These three classification tasks include primary classification, secondary classification, and tag classification.
We assign a primary-class label $y^1$, a secondary-class label $y^2$, and a tag label $y^3$ to each video in our system. 
The label $y^k$ ($k \in \{1,2,3\}$) is a binary vector in which if some elements are 1, the input video belongs to the classes corresponding to these elements.
$y^1$, $y^2$, and $y^3$ describe the input video from coarse to fine, we use them to define our loss function as:
\begin{equation}
\mathcal{L}=-\sum_{k=1}^{N=3}\sum_{j=1}^{M^k}y^k_j\log(p^k_j),
\label{fm:fm2}
\end{equation}
where $M^k$ is the length of the binary vector $y^k$. 
Note that we construct our loss function based on one input video and our loss function is a cross-entropy loss corresponding to three classification tasks.
We use Equation \ref{fm:fm2} to fine-tune the CNN module $f_C$ and the transformer module $f_T$ to ensure the deep joint representation $z$ is effective to represent and summarize the multimodal data from the input video.
The deep joint representation $z$ is concatenated with the tag representation $\overline{e}$ to represent a video for clustering. 
How to generate the tag representation is illustrated in Section \ref{sec:graph}.
Note that we do not use tags as one input modality for the multi-modal multi-task representation learning algorithm.
We empirically find that adding the tag representation into the MMoE network leads to this network focus on the tag representation while ignoring the deep joint representation from the cover image and title.
The reason is that the tag representation contains strong semantic information with human prior.
Using such joint representations for clustering leads to that topics lack the semantic from the cover image and title.

\renewcommand{\tabcolsep}{.8pt}
\def\swone{0.24\linewidth}
\begin{figure*}[t]
\begin{center}
\begin{tabular}{cccc}
\includegraphics[width=\swone]{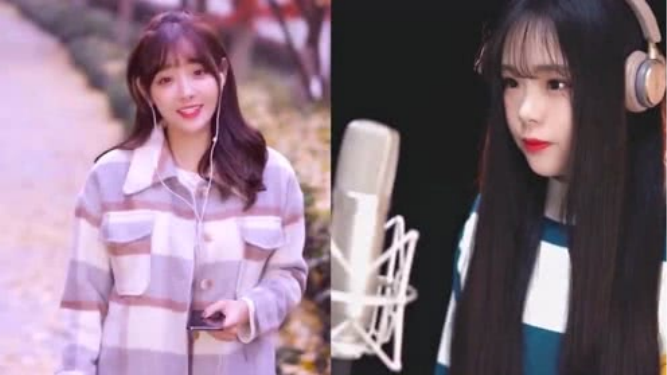}&
\includegraphics[width=\swone]{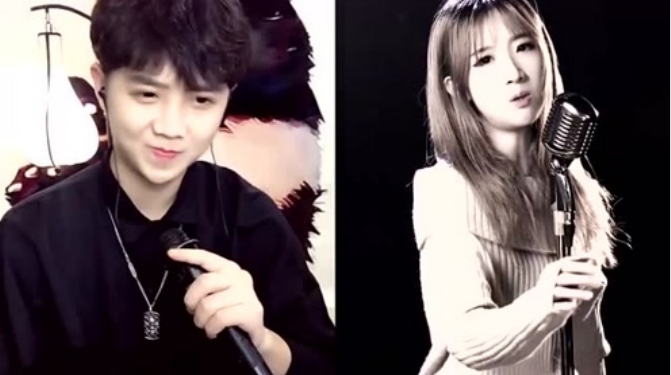}&
\includegraphics[width=\swone]{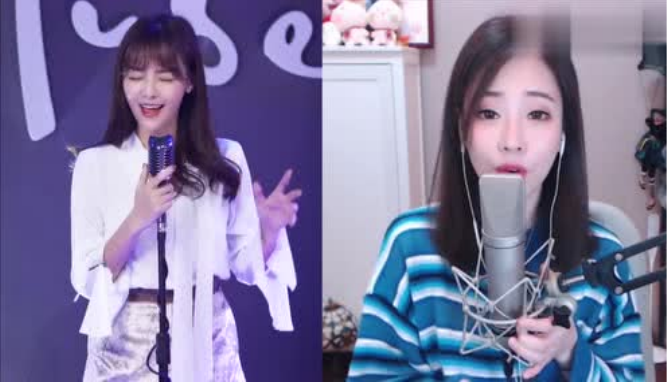}&
\includegraphics[width=\swone]{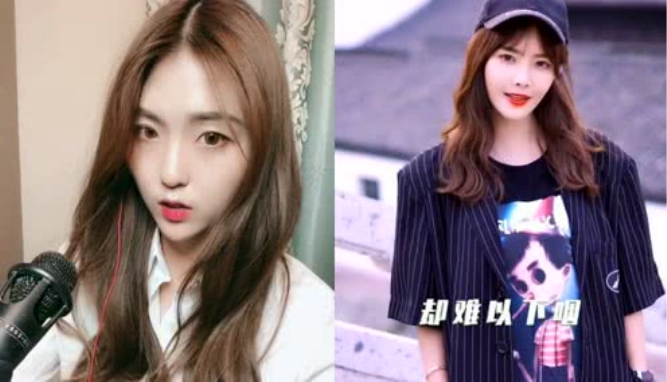}\\
\includegraphics[width=\swone]{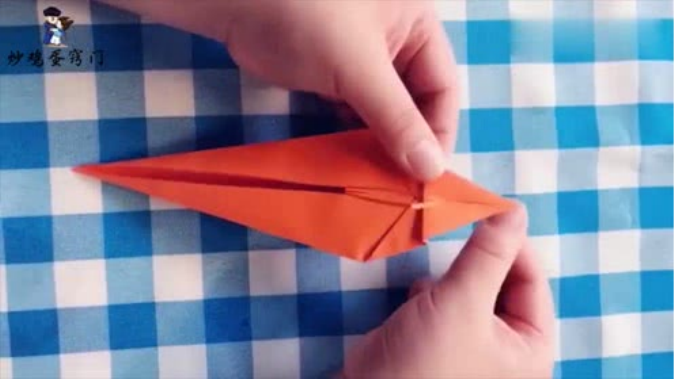}&
\includegraphics[width=\swone]{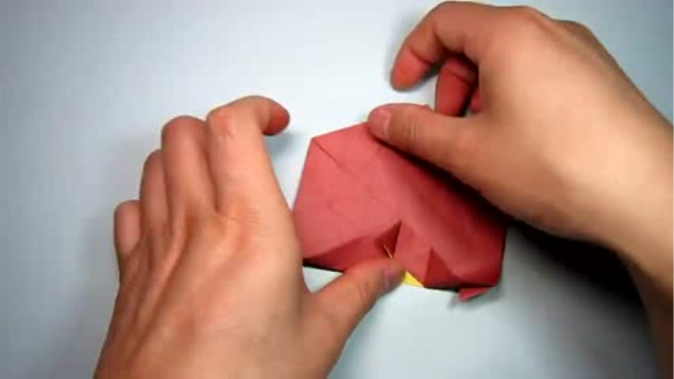}&
\includegraphics[width=\swone]{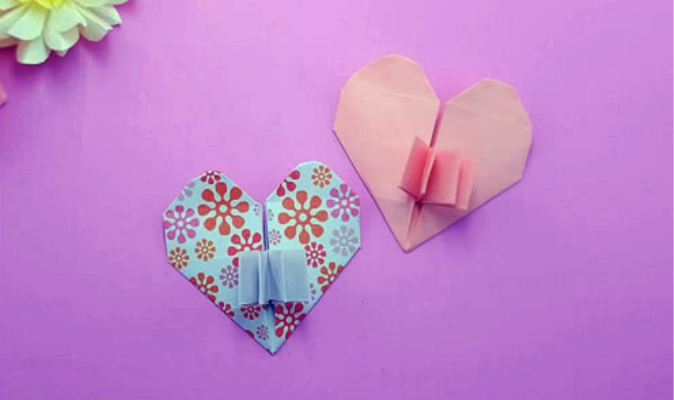}&
\includegraphics[width=\swone]{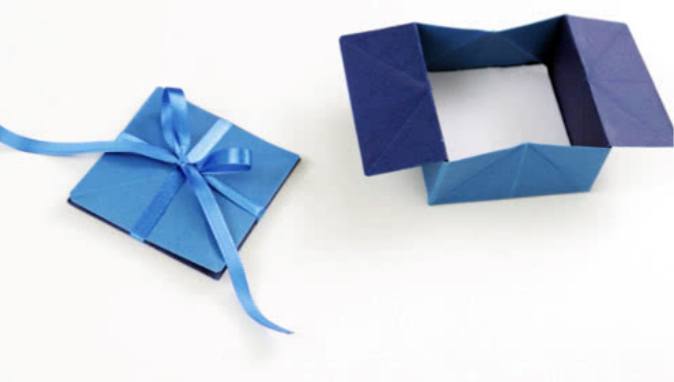}\\
\includegraphics[width=\swone]{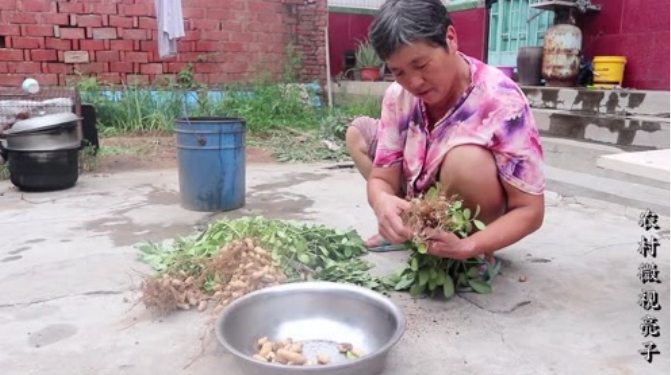}&
\includegraphics[width=\swone]{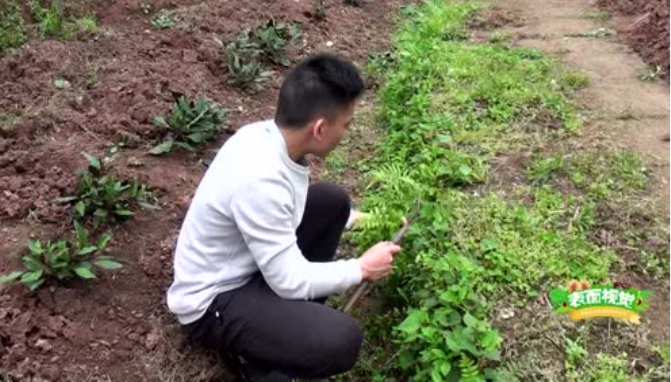}&
\includegraphics[width=\swone]{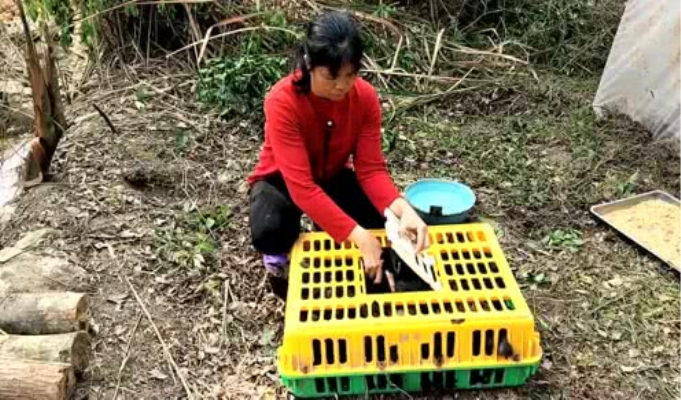}&
\includegraphics[width=\swone]{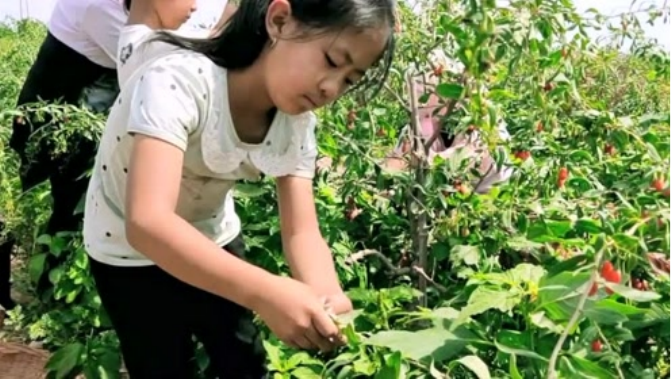}
\end{tabular}
\end{center}
\caption{Visualization of cover images from the videos in three topic queues. The cover images in each row are from the same topic queue.}
\label{fig:fig4}
\end{figure*}
\subsubsection{Graph embedding}
\label{sec:graph}
There are thousands of tags in our system while numerous tags contain similar semantic. 
Thus, it is not suitable to use one-hot coding to represent the tags of a video.
We propose a graph embedding approach to generate one tag representation for each video in our system.
%
%
%
In this approach, we set a tag as a node and convert all tags in our system into a completely connected graph.
We set the value of the edge linking any two tag nodes as the number of the videos which contain these two tags.
We remove the edges which value is 0 and generate the probability of each remaining edge as:
\begin{equation}
\begin{split}
p_\varepsilon=&\frac{x_\varepsilon}{\sum_{j=1}^E x_j},\\
\varepsilon \in& \{1,\cdots,E\},
\label{fm:fm3}
\end{split}
\end{equation}
where $E$ is the number of edges, $x_\varepsilon$ is the value of $\varepsilon$-th edge and $p_\varepsilon$ is the probability of this edge.

After we obtain the probability graph, we perform the deep walk \cite{perozzi2014deepwalk} and skip-gram algorithms to generate a tag embedding list that contains each tag and its corresponding embedding. 
Specifically, we simulate a random walk beginning from any tag node $t_1$ to generate a tag sequence with the length $l$.
The tag node $t_n$ $(n\in\{1,\cdots,l\})$ in this sequence is taken by sampling based on the probabilities of edges which linking the tag node $t_{n-1}$.
We collect a set of tag sequences and use the skip-gram approach to handle these tag sequences to generate the tag embedding list which can be denoted as $\{t_1:e_1,\cdots,t_T:e_T\}$, where $T$ is the number of tags in our system, $e_j$ $(j\in(1,\cdots,T))$ is the embedding of the tag $t_j$. Note that $T$ is the same as $M^3$ in Equation \ref{fm:fm2}, we use $T$ here for presentation clarity.

Given tags $\{t_1,\cdots,t_v\}$ of a video, we lookup the tag embedding list to generate tag representations $\{e_1,\cdots,e_v\}$. The tag representation of this video is denoted as:
\begin{equation}
\begin{split}
\overline{e}=\frac{1}{v}\sum_{j=1}^ve_j
\label{fm:fm4}
\end{split}
\end{equation}

\subsubsection{Clustering process}
We concatenate the tag representation $\overline{e}$ and the deep joint representation $z$ to be the final topic representation $a$ of a video. 
The L2 normalization method is used for $a$ to ensure that $\overline{e}$ and $z$ contribute the same amount for the distance metric in the clustering process. 
We use the K-means algorithm \cite{likas2003global} for clustering based on the topic representations from a large number of videos.
The $K$ clustering centers of these topic representations are treated as our topics.
We assign topic IDs for each video according to the cosine similarity distances between the topic representation $a$ of this video and the clustering centers considering the distance threshold $\tau_3$.
Each video in our system contains an average of 4 topics.

To present the effectiveness of our topic representation $a$ integrating multiple modalities, Figure \ref{fig:fig3} shows the distance ratios between average inter-class distances and average intra-class distances in 10 topic clustering results. 
We note that the distance ratios of our method using the cover image, title, and tags of each video are greater than that of the method using the cover image and title and that of the method only using the cover image.
This visualization demonstrates that our topic representation with multimodal information is more discriminative for the topic generation.


%
\subsection{Topic utilization}
After we use the MTL algorithm to assign topics for each video, we use the topics in the candidate generation module and the ranking module of our recommendation system. In this following, we introduce how to use the topics in both modules.
\subsubsection{Candidate generation module}
The candidate generation module retrieves hundreds of videos as candidates for the subsequent ranking module.
We use topics to facilitate the candidate generation module to determine preference scope.
Specifically, we first set $K$ topic queues, then assign all videos in our system to these queues according to the topics of these videos.
In each topic queue, we delete the videos whose exposure times and click-through rate are less than the thresholds $\tau_1$ and $\tau_2$, respectively. 
In addition, we randomly remain some fresh videos in this queue.
These fresh videos account for 10\% of all remaining videos in this queue.
After that, we update the video rankings in each topic queue according to click-through rate in real-time.
Figure \ref{fig:fig4}  shows cover images of the videos in three topic queues.
When recommending videos to a user, we count the number of times each topic in our system appears in the recent $N_1$ click videos of the user.
We select the top $N_2$ topics of the user and generate $N_3$ video candidates from the corresponding topic queues according to the ranking of the videos in these queues.
These video candidates are deliver to the subsequent ranking module.

\subsubsection{Ranking module}
The ranking module assigns scores to candidates from the candidate generation module.
A few video candidates with the highest scores are ranked by the scores and shown to the user.
We use topics to improve the ranking process for recommendation generation.
Our ranking module is based on the Wide\&Deep framework \cite{cheng2016wide}.
This framework transfer an ID feature to an embedding.
The embedding from a video and the mean embedding from the historical click videos of a user are combined to input the deep module of this framework.
The deep module of this framework uses multiple types of ID feature to generate the recommendation score of the video for the user.
Our topics serving as a type of ID features are integrated into this framework.

\section{Experiments}
In this section, we first present the implementation details, dataset, and evaluation metrics of our method. Then we conduct ablation studies to analyze the effect of of each module in our model on performance improvement. Finally, we offline and online evaluate our method on a large scale dataset and on the Kuaibao information streaming platform.
\subsection{Implementation details}
In the multimodal multitask learning method, we construct our CNN and transformer module based on the Inception-ResNet-v2 network \cite{szegedy2016inception} and the BERT network \cite{devlin2018bert}. 
The number $H$ of the expert networks in the MMoE is set as 12. 
The tasks in the MMoE include primary classification, secondary classification, and tag classification. Thus $N$ in Equation \ref{fm:fm1} is 3. 
$M^1$, $M^2$, and $M^3$ in Equation \ref{fm:fm2} are 50, 367, and 12000 which are the number of primary-classes, the number of secondary-classes, and the number of tags in our system.
We use the adaptive moment estimation optimizer to train the network using 200 iterations.
The learning rate is 0.001 and the batch size is set as 300.
The length $l$ of a tag sequence in the graph embedding method is 100. We set the number of clustering centers $K$ as 6000 and the cosine similarity distances threshold $\tau_3$ is 0.6.
In our recommendation system, the thresholds $\tau_1$ and $\tau_2$ are set as 10 and 0.07, respectively.  
We set $N_1$, $N_2$, and $N_3$ as 128, 4, and 600.
The proposed MTL method is executed on a PC with Intel(R) Xeon(R) CPU E5-2680 v4 @ 2.40GHz CPU and 8 Tesla P40 GPU.

\subsection{Datasets  and evaluation metrics}
Our experimental data is from the Kuaibao information streaming platform. 
Specifically, We construct a video content dataset and a click dataset for model learning and ablation studies.
The video content dataset contains about 3 million samples. 
Each sample consists of the title, cover image, primary-class label, secondary-class label, and tags. 
There are 50 primary-classes, 367 secondary-classes, and 12000 video tags in this video content dataset.
We use 90\% of the samples in the video content dataset for the proposed topic learning and the remaining 10\% of the samples are used to evaluate the representation ability of features generated by different models.
The metrics for the representation ability evaluation are Adjusted Rand Index (ARI) and Normalized Mutual Information (NMI). 
The click dataset contains about 14 million samples which are produced in 5 days.
Each sample is a click that corresponds to a user and a video.
There are about 1.5 million users and 800 thousand videos in this click dataset.
We use the samples from the first 4 days to train recommendation models, and then use the samples from the last day to offline evaluate the effect of topics generated by our algorithm on recommendation performance improvement.
The metric for offline evaluation is Area Under ROC Curve (AUC) which is widely adopted in recommendation systems \cite{chen2016deep,mo2015image,zhou2018deep}.

For online evaluation, we perform an online A/B test on the Kuaibao.
The metrics for the online A/B test include User Video Duration (UVD), User Average Video Duration (UAVD), Average Video Duration(AVD), and Average Refresh Times (ART).
UVD measures the total watching time for all users.
UAVD can be denoted as $\frac{\text{UVD}}{\text{The number of users}}$.
AVD measures the average watching time for each video.
ART is the average number of times for each user to refresh the recommendation page.

\renewcommand{\tabcolsep}{3pt}
\begin{table}
	\small
	\begin{center}
		\caption{Representation comparisons with different models on the video content dataset. The results are presented in terms of the Adjusted Rand Index (ARI) and Normalized Mutual Information (NMI).}
		\label{table:representation}
		\begin{tabular}{lp{1cm}<{\centering}p{1cm}<{\centering}p{1.1cm}<{\centering}p{1.1cm}<{\centering}p{1.1cm}<{\centering}p{1.1cm}<{\centering}p{1.1cm}<{\centering}p{1.1cm}<{\centering}p{1.1cm}<{\centering}p{1.1cm}<{\centering}p{1.1cm}<{\centering}p{1.1cm}<{\centering}}
			\hline\noalign{\smallskip}
			&&IRv2&BERT&MMoE&MMoE\\	
			&&&&&+Tag\\		
			\noalign{\smallskip}			
			\hline
			\noalign{\smallskip}
			\multicolumn{1}{l}{\multirow{2}*{$\text{ARI}(\permil)\uparrow$}}&PC&69.3&1.0&171.3&\bf237.1 \\
			~&SC&17.6&2.4&31.3&\bf53.0\\
			\hline
			\noalign{\smallskip}
			\multicolumn{1}{l}{\multirow{2}*{$\text{NMI}(\permil)\uparrow$}}&PC&186.0&32.7&309.2&\bf427.2\\
			~&SC&227.3&71.1&314.4&\bf455.2\\
			\hline
		\end{tabular}
	\end{center}
	
\end{table}

\subsection{Ablation studies}
In this section, we construct ablation studies from two perspectives:
1) We compare the representation ability of features generated by different models to demonstrate the effectiveness of our multi-task learning and graph embedding methods.
2) We add semantic topic features into existing recommendation algorithms to investigate whether the topics can improve performance generically.

Table \ref{table:representation} shows the representation ability evaluation.
We transfer the representation ability comparisons to clustering performance comparisons.
Specifically, we use the K-means algorithm \cite{likas2003global} to cluster the samples from the video content dataset.
The value of K in this algorithm is set as the number of primary-classes (PC) or secondary-classes (SC) in this dataset.
We use the ARI and NMI metrics to measure the clustering performance corresponding to the distributions of the primary-classes or the secondary-classes.
As we use the same clustering algorithm to cluster sample features from different models, the clustering results reflect the representation ability of different features and evaluate the performance of different feature extractors.
We firstly set the Inception-ResNet-v2 (IRv2) network \cite{szegedy2016inception} and the BERT network \cite{devlin2018bert} as our baselines.
These two networks extract features from cover images and from titles to represent samples, respectively.
Then we use the MMOE framework with multi-task learning to end-to-end train these two baselines, the joint representations from these two networks are used to represent samples.
Finally, we combine the joint representations with the tag features generated by our graph embedding method to represent samples.
Table \ref{table:representation} shows that our multi-task learning method significantly improves the clustering performance. Combining the tag features achieves further improvements.
We attribute the performance improvements to that our proposed algorithm is effective to fuse multimodal information to represent video content.
\renewcommand{\tabcolsep}{.8pt}
\def\swone{1\linewidth}
\begin{figure}[t]
\begin{center}
\begin{tabular}{c}
\includegraphics[width=\swone]{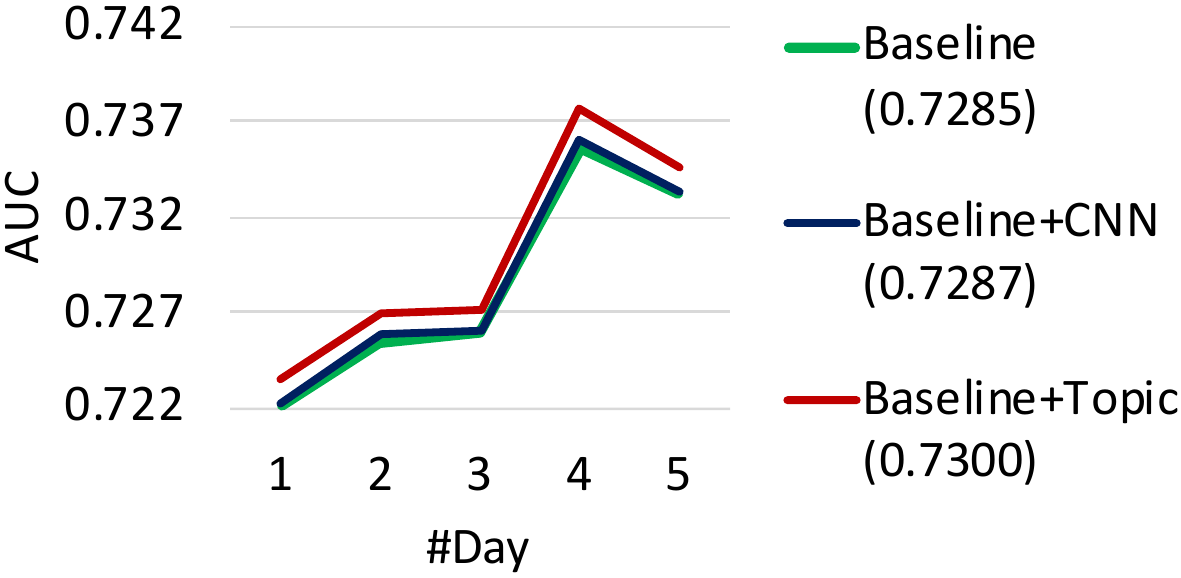}
\end{tabular}
\end{center}
\caption{AUC comparisons in offline evaluation using click data from 5 consecutive days. The legend contains the average AUC results.}
\label{fig:offlineAUC}
\end{figure}

To present the generalization ability of the semantic topic features, Table \ref{table:recom} shows the comparisons between classical recommendation algorithms with and without the semantic topic features. 
These algorithms include the Logistic Regression (LR), Embedding\&MLP (EMLP), and DCN \cite{wang2017deep}.  The AUC metric is used to evaluate recommendation performance on the click dataset.
Table \ref{table:recom} shows that existing recommendation algorithms with the topic features consistently achieve superior AUC results against these methods without the topic features, which reflects the generalization and improvements brought by the proposed semantic topics.

\renewcommand{\tabcolsep}{3pt}
\begin{table}[h]
	\small
	\begin{center}
		\caption{Evaluation for recommendation performance of classical recommendation algorithms with (w/) and without (w/o) topic features on the click dataset. The metric is the Area Under ROC Curve (AUC). }
		\label{table:recom}
		\begin{tabular}{lp{1cm}<{\centering}p{1cm}<{\centering}p{1.5cm}<{\centering}p{1.5cm}<{\centering}p{1.5cm}<{\centering}}
			\hline\noalign{\smallskip}
			&&LR&EMLP&DCN\\	
			\noalign{\smallskip}			
			\hline
			\noalign{\smallskip}
			\multicolumn{1}{l}{\multirow{2}*{$\text{AUC}(\%)\uparrow$}}&w/o&46.23&80.40&83.01\\
			~&w/&\bf46.91&\bf81.17&\bf83.76\\	
			\hline
		\end{tabular}
	\end{center}
	
\end{table}

\subsection{Offline evaluation}
In the offline evaluation, we use historical data from the Kuaibao to simulate the online process.
Specifically, the process includes 3 steps:
(1) We initialize the recommendation algorithms using the click dataset.
(2) We use the recommendation algorithms to predict clicks in one day.
(3) We collect the predicted clicks to update recommendation algorithms online.
After that, we repeat the (2) and (3) steps.
We collect the predicted clicks and real clicks in 5 consecutive days to perform the offline evaluation.
The evaluation results are presented in terms of the AUC metric, the memory cost, and the time cost.

In this evaluation, the baseline method is the same as the production model in the Kuaibao. 
We respectively add the CNN features from the cover image and the semantic topic features into the baseline method to perform comparisons.
 These two configurations are denoted as Baseline+CNN and Baseline+Topic.
 Figure \ref{fig:offlineAUC} shows that the CNN features slightly improve the AUC  results of the baseline, while our semantic topic features improve the baseline method by a large margin. 
 The reason is that directly using CNN features in the recommendation systems can not integrate the global information related to the video pool of the recommendation systems. 
 The visual content information is not enough to achieve significant improvements. 
 Our semantic topic features are offline learned on a large scale dataset from the video pool. 
 The high-level semantic information encoded in the topic features effectively help recommendation systems to understand the content preferences of users.
 In addition, Figure \ref{fig:offline_cost} shows the computational cost comparisons. 
 We deliver one sample to the three configurations to compute the time and memory costs.
 The increments of the time and memory costs of the Baseline+CNN method are about 3 times greater than that of the Baseline+Topic approach.
As CNN features have to remain high dimensionality (128 dimensionalities in the Baseline+CNN method) to integrate visual content information, it is inevitable to increase the computational cost.
Our semantic topic features with low dimensionality (8 dimensionalities in the Baseline+Topic) effectively reduce the online computational cost brought by the CNN features.

\renewcommand{\tabcolsep}{.8pt}
\def\swone{0.5\linewidth}
\begin{figure}[t]
\begin{center}
\begin{tabular}{cc}
\includegraphics[width=\swone]{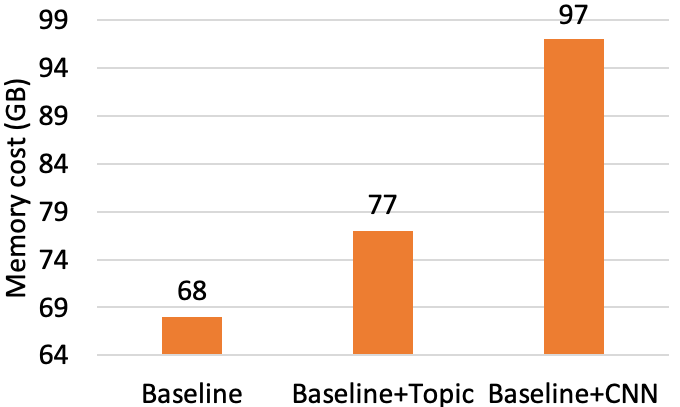}&
\includegraphics[width=\swone]{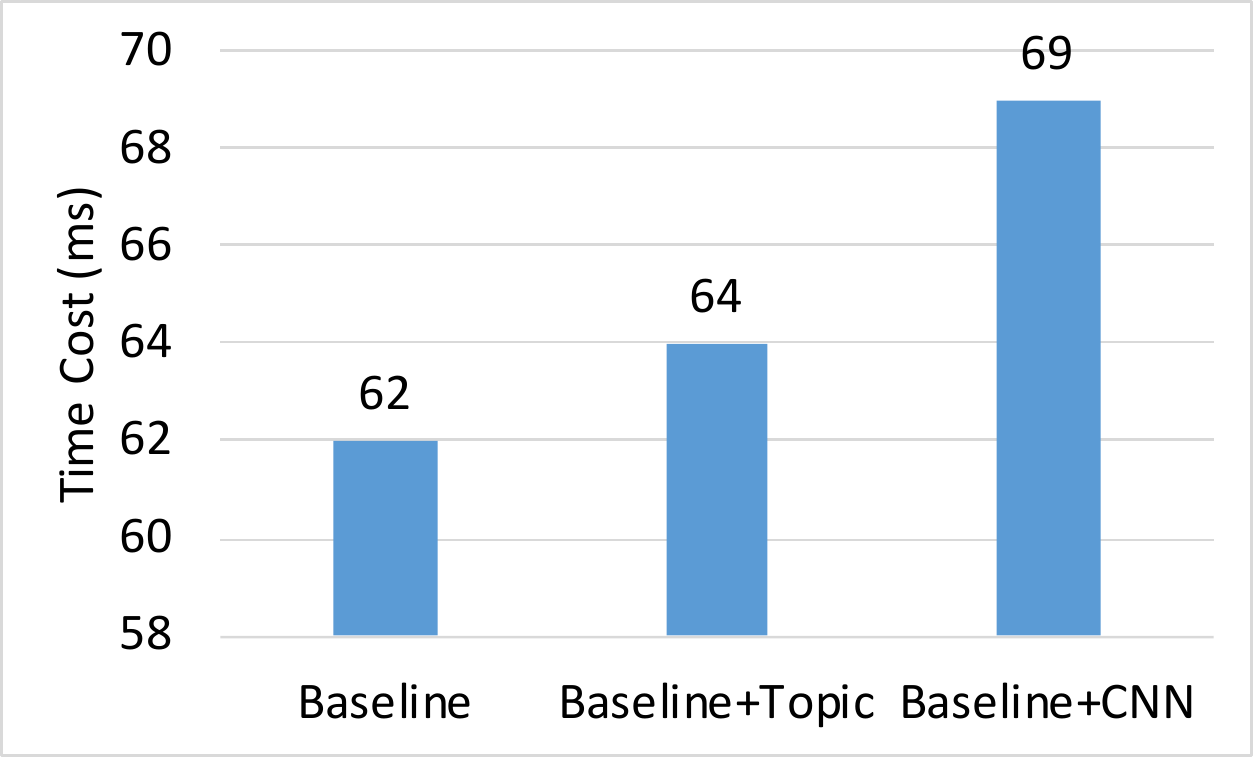}\\
(a)&(b)
\end{tabular}
\end{center}
\caption{Computational cost comparisons in offline evaluation using click data from 5 consecutive days. }
\label{fig:offline_cost}
\end{figure}

\renewcommand{\tabcolsep}{3pt}
\begin{table}[h]
	\small
	\begin{center}
		\caption{Relative increments of adding topics into the baseline method in the online A/B test. }
		\label{table:online}
		\begin{tabular}{lp{3cm}<{\centering}p{3cm}<{\centering}}
			\hline\noalign{\smallskip}
			&Recall+Topic&Baseline+Topic\\	
			\noalign{\smallskip}			
			\hline
			\noalign{\smallskip}
			UVD$(\%)\uparrow$&+0.57&+0.82\\
			\noalign{\smallskip}
			UAVD$(\%)\uparrow$&+1.51&+3.01\\
			\noalign{\smallskip}
			AVD$(\%)\uparrow$&+0.11&+1.01\\
			\noalign{\smallskip}
			ART$(\%)\uparrow$&+0.56&+0.71\\
			\noalign{\smallskip}
			\hline
			
		\end{tabular}
	\end{center}
	\vspace{-0.3cm}
	
\end{table}

\subsection{Online evaluation}
We perform an A/B test in the online evaluation.
The process of the A/B test consists of 3 steps.
Firstly, we divide 300, 000 users into 8 groups on average.
Then, we measure the activity of each group for 7 days in terms of the UVD, UAVD, AVD, and ART metrics and we select two groups with similar activities as the experiment group and the control group.
Finally, we execute our proposed approach and the baseline method in the experiment group and the control group respectively for 7 days.
The UVD, UAVD, AVD, and ART metrics are used to evaluate test results.

In the online A/B test, we set the baseline method as the production model in the Kuaibao.
We gradually add the topics generated by our proposed algorithm into the candidate generation (i.e., Recall) and ranking modules of the video recommendation model in the Kuaibao.
We denote the two configurations as Recall+Topic and Baseline+Topic.
Table \ref{table:online} shows that when we use the topic in the candidate generation module (i.e., Recall+Topic), the recommendation performance is improved slightly.
However, when we further add the topic features into the rank module (i.e., Baseline+Topic), we achieve significant performance improvements.
Our proposed topics facilitate these two modules to advance existing recommendation systems.

\section{Conclusion}
In this paper, we propose a multimodal topic learning algorithm to generate video topics offline.
We use the semantic topics as video content features to facilitate the candidate generation and ranking modules in our recommendation system.
The semantic topics help our recommendation system to determine preference scope and generate recommendations.
Online and offline experimental results demonstrate the effectiveness of our proposed algorithm.






{\small
\bibliographystyle{ieee}
\bibliography{egbib}
}

\end{document}